\documentclass[3p,times]{elsarticle}

\usepackage{ecrc}

\volume{00}

\firstpage{1}

\journalname{Nuclear Physics A}

\jid{nupha}

\usepackage{amssymb}
\usepackage[figuresright]{rotating}

\begin{document}

\begin{frontmatter}


%% \author{Name\corref{}\fnref{}}
\author{Elisa Incani for the ALICE collaboration}
\ead{elisa.incani@cern.ch}

\dochead{}

\title{Light vector meson production at the LHC with the ALICE detector}


%%\author{}

\address{Universit\`a  and INFN Cagliari, Complesso Universitario di Monserrato, 09042 Monserrato (CA), Italy}

\begin{abstract}
The measurement of light vector meson production ($\rho$, $\omega$, $\phi$) in pp
collisions provides insight into soft Quantum Chromodynamics (QCD)
processes in the LHC energy range. Calculations in this regime are based
on QCD inspired phenomenological models that must be tuned to the data.
Moreover, light vector meson production provides a reference for
high-energy heavy-ion collisions. 

A measurement of the $\phi$ and $\omega$ differential cross sections was performed by the ALICE experiment  
in pp collisions at $\sqrt{s}=$7~TeV and of the $\phi$ cross section
in pp collisions at $\sqrt{s}=$2.76~TeV through their decay to muon pairs and in the rapidity interval $2.5 < y < 4$.
\end{abstract}

%\begin{keyword}


%\end{keyword}

\end{frontmatter}

% \linenumbers

%\section{ }
%\label{}

Vector meson production in pp collisions can be used to tune particle production models in the
unexplored LHC energy range and they are key probes of the hot and dense state of strongly interacting matter produced in heavy ion collisions. 
Their dileptonic decay channel is particularly suitable for these studies, since dileptons have negligible final state 
interactions in QCD matter.

The ALICE experiment at the LHC can access vector mesons produced at forward rapidity through. 
The detector is fully described in [\citenum{ALICE}]. 
In this paper we report results from the analysis of the data collected 
during the 2010 pp run at $\sqrt{s}=$7~TeV (already published in [\citenum{LMR7TeV}]) and during the 2011 pp run at
$\sqrt{s}=$2.76~TeV in the rapidity range $2.5<y<4$ through their decays to muons pairs.

The invariant mass spectrum for opposite sign muon pairs was measured for $1<p_T<5$~GeV/$c$ for the data at 7~TeV and for 
$1<p_T<4$~GeV/$c$ for the data at 2.76~TeV.
Muon tracks were selected asking that the tracks reconstructed in the tracking stations matched the ones in the trigger chambers
 and that their pseudorapidity was in the range $2.5 <\eta_\mu<4$. Muon pairs were selected requiring that the dimuon rapidity was
inside the interval $2.5<y_{\mu\mu}<4$.
The combinatorial background in the opposite sign dimuon spectrum was subtracted using an event mixing technique.

The resulting mass spectra were described as a superposition of light meson decays to muon pairs and charm and beauty semi-muonic decays. 
From the fit of the dimuon invariant mass spectra, it was possible to extract the number of $\phi$ ($N_\phi=3200\pm150$ at 7~TeV and $N_\phi=350\pm40$ at 2.76~TeV) 
and $\rho+\omega$ ($N_{\rho+\omega}=6830\pm150$ at 7~TeV and $N_{\rho+\omega}=801\pm43$ at 2.76~TeV).
\begin{figure}[h!]
\centering
\includegraphics[width=0.45\textwidth]{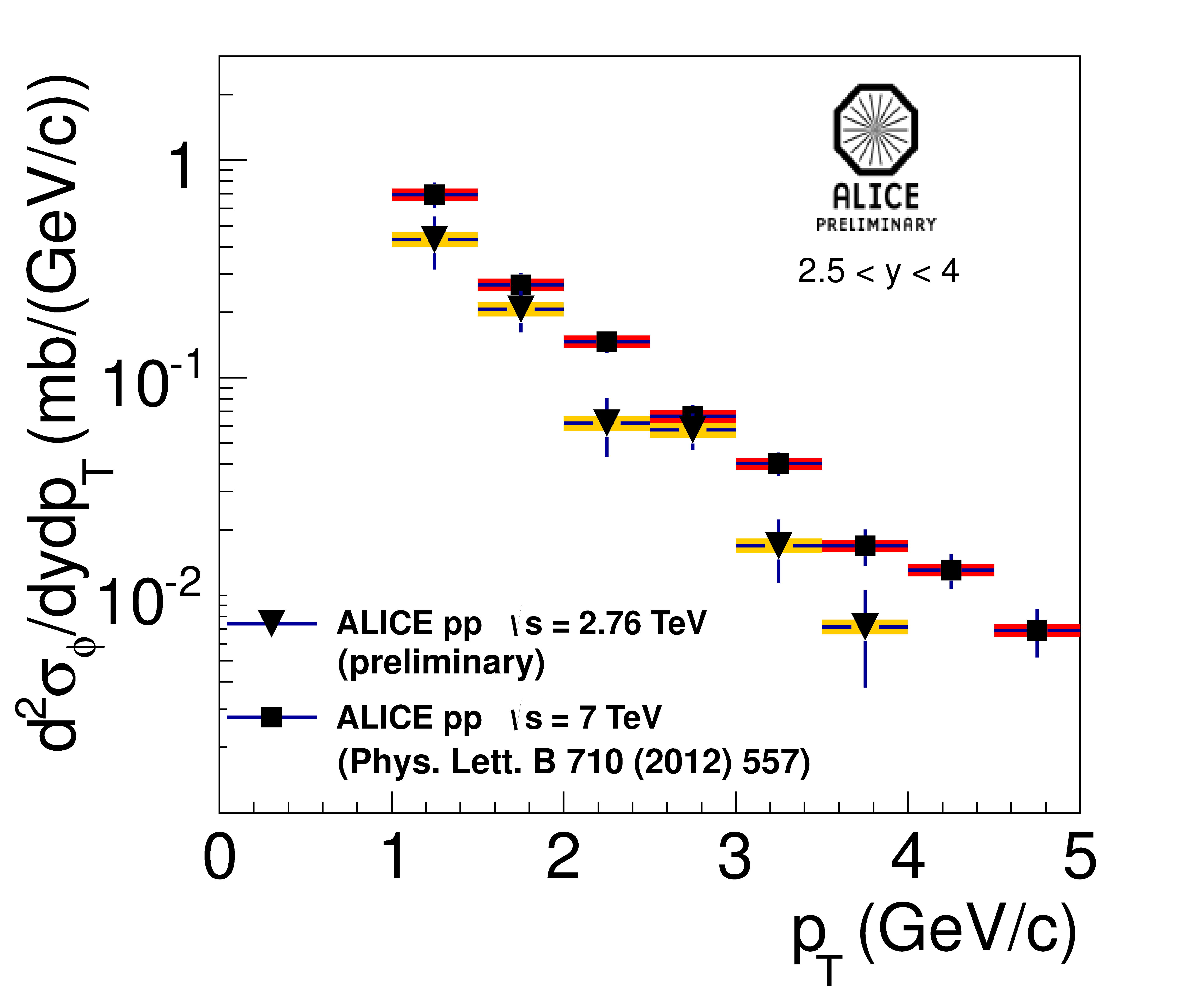}
\includegraphics[width=0.39\textwidth]{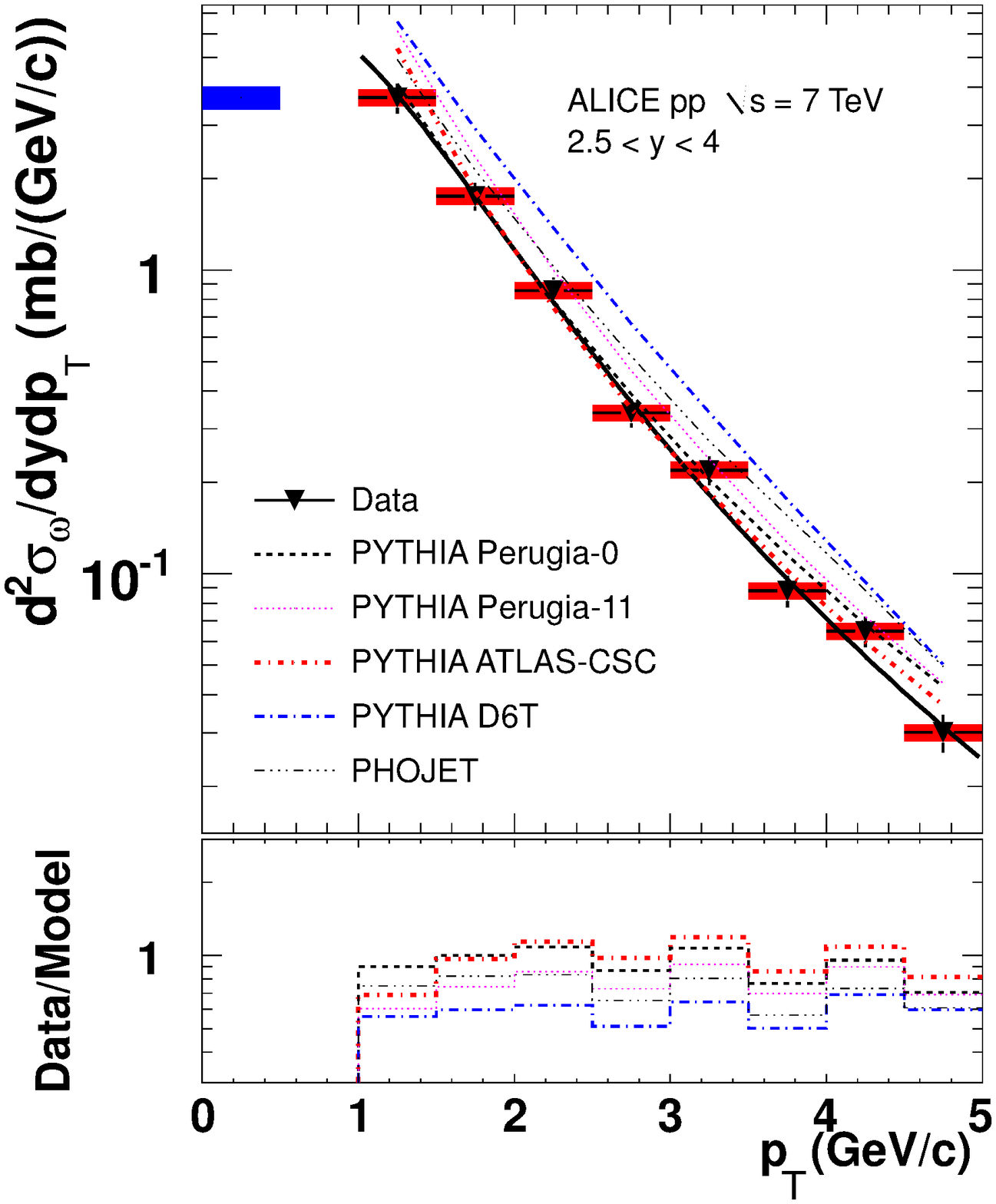}
\caption{Differential cross sections $d\sigma^2 _\phi/dydp_T$ for data at 7 TeV and 2.76 TeV (left) and differential cross section $d\sigma^2 _\omega/dydp_T$ for data at 7 TeV (right).}
\label{fig:phi}
\end{figure}
The cross sections were extracted from a sample corresponding to an integrated luminosity of 
$\mathrm{L}_{int}$=55.7~nb$^{-1}$ for the data at 7~TeV and  $\mathrm{L}_{int}$=17.6~nb$^{-1}$ for the data at 
2.76~TeV, through the formula 
$\sigma_\phi=\frac{N_\phi^c}{BR(\phi\rightarrow\mu^+\mu^-)}\frac{\sigma_{MB}}{N_{MB}}\frac{N^{MB}_\mu}{N^{\mu-MB}_\mu}$,
where $N^c_\phi$ is the measured number of $\phi$ mesons corrected for the efficiency and acceptance, 
$N_{MB}$ is the number of minimum bias collisions, $\sigma_{MB}$
is the minimum bias cross section  (at 7~TeV the value is 62.3 mb and at 2.76~TeV it is 54.8 mb \cite{VM})
, $N_\mu^{MB}$ is the number of single muons in the region 
$2.5<y_\mu<4$, $p_{T\mu}>1$~GeV/c 
collected with the minimum bias trigger, and $N_\mu^{\mu-MB}$ is the number of muons in the same region collected with
the muon trigger. 
The $\phi$ cross section value in pp collisions at $\sqrt{s}=7$~TeV is 
$\sigma_\phi(1<p_T<5\ \textrm{GeV}/$c$, 2.5<y<4)=0.940\pm0.084(stat.)\pm0.076(syst.)$~mb 
and in pp collisions at $\sqrt{s}=2.76$~TeV is 
$\sigma_\phi(1<p_T<4\ \textrm{GeV}/$c$, 2.5<y<4)=0.587\pm0.070(stat.)\pm0.045(syst.)$~mb.
In Fig. \ref{fig:phi}, left side, the $p_T$-differential cross section $d\sigma^2 _\phi/dydp_T$ is shown for the data at 7~TeV (black squares) and at~2.76 TeV 
(black triangles).
Data at 2.76 TeV were compared with some commonly used models: PHOJET [\citenum{PHOJET}, \citenum{PHOJET2}] and PYTHIA [\citenum{PYTHIA}] (Perugia-0 [\citenum{Perugia-0}], Perugia-11 [\citenum{Perugia-11}],
ATLAS-CSC [\citenum{ATLAS}] and D6T [\citenum{D6T}] tunes), for which the cross sections are $\sigma_\phi=0.487$~mb, $\sigma_\phi=0.275$~mb,  $\sigma_\phi=0.293$~mb,
 $\sigma_\phi=0.464$~mb and $\sigma_\phi=0.625$~mb respectively. 
PHOJET and PYTHIA with ATLAS-CSC and D6T tunes reproduce the $\sigma_\phi$, while Perugia-0 and Perugia-11 underestimate the data.
Similar results are observed at 7~TeV~[\citenum{LMR7TeV}]. 

The data at 7~TeV have been compared also with the measurements in kaon pairs performed by the LHCb Collaboration in a similar rapidity range [\citenum{lhcb}]: the shapes 
are similar and the rescaling of the LHCb cross section for $p_T>1$~GeV/$c$ and for $2.5<y<4$ gives $\sigma_\phi=1.07\pm0.15$ (full error) mb, which is in agreement with the ALICE 
measurement in the same phase space domain.

In order to extract the $\omega$ cross section at 7~TeV, the $\rho$ and $\omega$ contributions must be disentangled, leaving the $\rho$ normalization as an
additional free parameter in the fit to the dimuon mass spectrum.
The result of the fit gives the ratio $\frac{\sigma_\rho}{\sigma_\omega}=1.15\pm0.20 (stat.)\pm0.12(syst)$
leading to
$\sigma_\omega(1<p_T<5)\ \textrm{GeV}/$c$, 2.5<y<4)=5.28\pm0.54(stat.)\pm0.49(syst.)$~mb.
In Fig. \ref{fig:phi}, right side, the $p_T$-differential cross section $d\sigma^2 _\omega/dydp_T$ is shown,
the fit to the differential cross section with a power law function gives $p_0 = 1.44 \pm 0.09$~GeV/$c$ and $n = 3.2\pm0.1$.
Comparison with some commonly used tunes shows that they overestimate $\sigma_\omega$ except for PYTHIA Perugia-0.

For Pb-Pb collisions the analysis for the extraction of the nuclear modification factor is in progress.

\vspace{-0.3cm}


%% References

\bibliographystyle{elsarticle-num}
\bibliography{bib}



%% \begin{thebibliography}{00}


%  \bibitem[K. Aamodt et al. (ALICE collaboration), J. Instrum 3, (2008) S08002]{ALICE}
%  \bibitem[K. Aamodt et al. (ALICE collaboration), J. Instrum 3, (2008) S08002]{VM}
% \bibitem[K. Aamodt et al. (ALICE collaboration), J. Instrum 3, (2008) S08002]{LMR7TeV}
% \bibitem[K. Aamodt et al. (ALICE collaboration), J. Instrum 3, (2008) S08002]{PHOJET}
% \bibitem[K. Aamodt et al. (ALICE collaboration), J. Instrum 3, (2008) S08002]{PHOJET2}
% \bibitem[K. Aamodt et al. (ALICE collaboration), J. Instrum 3, (2008) S08002]{PYTHIA}
% \bibitem[K. Aamodt et al. (ALICE collaboration), J. Instrum 3, (2008) S08002]{Perugia-0}
% \bibitem[K. Aamodt et al. (ALICE collaboration), J. Instrum 3, (2008) S08002]{Perugia-11}
% \bibitem[K. Aamodt et al. (ALICE collaboration), J. Instrum 3, (2008) S08002]{ATLAS}
% \bibitem[K. Aamodt et al. (ALICE collaboration), J. Instrum 3, (2008) S08002]{D6T}
% \bibitem[K. Aamodt et al. (ALICE collaboration), J. Instrum 3, (2008) S08002]{lhcb}

%% \end{thebibliography}

\end{document}